*"It is a riddle, wrapped in a mystery, inside an enigma; but perhaps there is a key…"*
*Winston Churchill*

# How Hidden Can Be Even More Hidden?

Wojciech Frączek, Wojciech Mazurczyk, Krzysztof Szczypiorski
Institute of Telecommunications
Warsaw University of Technology
Warsaw, Poland
e-mail: wfraczek@gmail.com,{wmazurczyk, ksz}@tele.pw.edu.pl

*Abstract* — The paper presents Deep Hiding Techniques (DHTs) that define general techniques that can be applied to every network steganography method to improve its undetectability and make steganogram extraction harder to perform. We define five groups of techniques that can make steganogram less susceptible to detection and extraction. For each of the presented group, examples of the usage are provided based on existing network steganography methods. To authors' best knowledge presented approach is the first attempt in the state of the art to systematically describe general solutions that can make steganographic communication more hidden and steganogram extraction harder to perform.

## I. INTRODUCTION

All of the information hiding techniques that may be used to exchange secret data (steganograms) in telecommunication networks are called network steganography. The term was originally introduced by Szczypiorski in 2003 [1]. To perform hidden communication network steganography utilizes as a carrier for steganograms network protocols and/or relationships between them [2]. It must be emphasized that for the third party observer who is not aware of the steganographic procedure steganograms exchange remains hidden. This is possible because inserting hidden data into chosen carrier must remain unnoticeable for users not involved into steganographic communication.

For each network steganography method there is always a trade-off necessary between maximizing steganographic bandwidth (how much data we are able to send using this particular method) and still remaining undetected. However, user can utilise a method naively and sends as much secret data as is possible but it simultaneously raises a risk of disclosure. So he/she must purposely resign from some fraction of the steganographic bandwidth in order to achieve undetectability. Such user behaviour raises the following questions: what else can user do to improve stealth of his/her secret communication? Does it all depend on the functioning of the particular steganographic method?

In this paper we introduce Deep Hiding Techniques (DHTs) which we define as general techniques that are able to make usual network steganography communication even more hidden. Thus such method is then harder to detect and the steganogram carried is less susceptible to extraction. Based on this concept we will show what measures can be taken for chosen steganographic method to make it harder to detect than without applying DHTs.

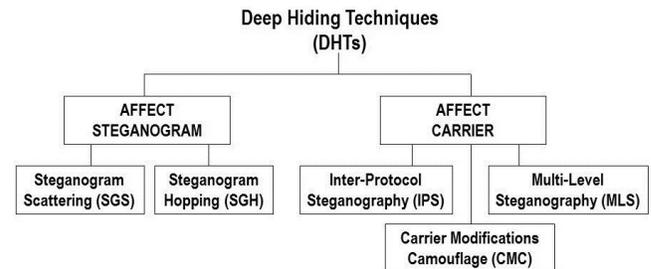

**Fig. 1 Deep Hiding Techniques classification**

We define five types of the Deep Hiding Techniques (presented classification (Fig. 1) is based on what particular DHT affect – a steganogram or a carrier):

- **Steganogram Scattering** (SGS) that utilizes different techniques which involve distributed sending of divided steganogram.
- **Steganogram Hopping** (SGH), which is based on the periodical change of the steganographic method during single hidden connection thus influencing the localisation of the steganogram.
- **Carrier Modifications Camouflage** (CMC), which focuses on masking insertion of the steganogram into hidden data carrier.
- **Inter-Protocol Steganography** (IPS), which uses relationships between two or more different network protocols to enable secret communication and make it harder to detect.
- **Multi-Level Steganography** (MLS), which allows to utilize functioning of the existing steganographic method (upper-level method), to create another one (lower-level method). The lower-level method completely relies on upper-level one.

The comparison between DHTs and typical network steganography method features is presented in Fig. 2.

The abovementioned techniques will be presented in detail throughout the rest of the paper which is structured as follows. Section 2 describes Steganogram Scattering techniques. Section 3 provides presentation of the Steganogram Hopping. Section 4 discusses Carrier Modifications Camouflage approach. Section 5 and 6 introduces Inter-Protocol and Multi-Level Steganography, respectively. Finally, Section 7 concludes our work.

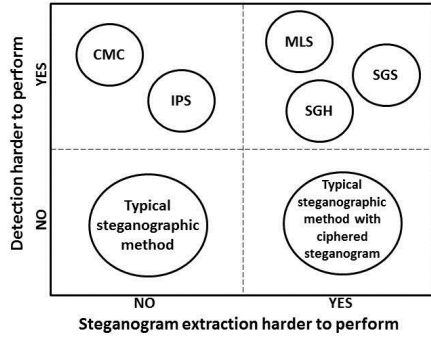

**Fig. 2 Deep Hiding Techniques vs. typical steganographic method features**

II. STEGANOGRAM SCATTERING (SGS)

*A. Description*

The idea of Steganogram Scattering techniques is to split steganogram into pieces and send it as separate messages. Each part may be transmitted using different steganographic method.

Currently, hidden communication methods are proposed that exploit division of the secret message into pieces. One of them is Collage [5], a method that hides parts of the steganogram in user-generated content on the Internet. However, to the best of authors' knowledge, scattering of the steganogram has not yet been considered as general tool that can be used with all known network steganography methods. We propose following classification of the SGS techniques (Fig. 2).

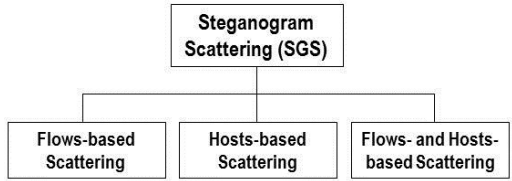

**Fig. 2 Steganogram Scattering (SGS) classification**

*1) Flows-based scattering*

Flows-based scattering takes advantage of the capability to set up many flows between two hosts. Each piece of the previously divided steganogram is sent using one of the available flows. In this case, flow means the possibility of the communication between two hosts, it is not confined to connection-oriented protocols (e.g. TCP) and flow can also be created using connectionless protocols (e.g. UDP). An idea of using many flows for purpose of network steganography method was introduced in Cloak method [6], but it was exploited as a mean to create new covert channel and limited only to TCP protocol.

An example of the flows-based SGS technique is presented in Fig. 3. User X wants to send hidden message to User Y. Sender decides to split steganogram into three pieces and use three flows: two TCP connections on port 80 (HTTP sessions with two pages of the website) and UDP communication on port 123 (NTP synchronization). Each part of the secret message is then sent using different flows and possibly different steganographic methods. User Y receives three separate chunks and combines them into original steganogram.

Presented example does not consider steganogram assembly process. However, it is important to determine order of the merging received fragments. For that purpose several possibilities can be utilized. The most naive one is to add information about a position in steganogram to each fragment. Another way is to make a position of the fragment in original steganogram depended on the time of sending. Consider two fragments – the one which sending was started earlier occupies position before another. For example, if a steganogram is divided into three parts and the first bit of fragment 3 is sent at 00:00:00, the first bit of the fragment 2 is sent at 00:00:01 and the first bit of the fragment 1 is sent at 00:00:02, it means that the first piece of the steganogram is fragment 3, the second one is fragment 2 and the third one is fragment 1. In other words, message can be correctly assembled if the order of the pieces in combined steganogram is: fragment 3, fragment 2 and fragment 1. It is worth noting that this method requires synchronization between sender and receiver. It is also possible to number flows and assign pieces of the steganogram to flows beforehand.

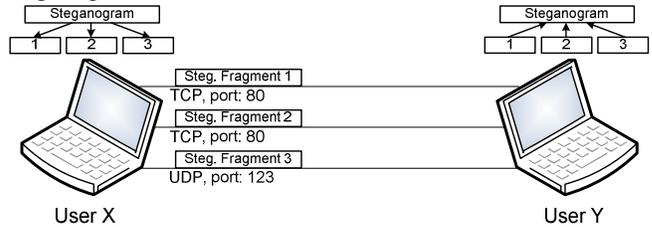

**Fig. 3 Flow-based Steganogram Scattering (SGS)**

*2) Hosts-based scattering*

Hosts-based scattering requires that a sender and/or receiver is controlling more than one physical host or other networking device. Pieces of the steganogram are hidden in different hosts-based overt channels (overt channels between two hosts). Two hosts-based overt channels are different if they have different senders, different receiver or both. The example of the hosts-based SGS is presented in Fig. 4. User X (*n* hosts) wants to send secret message to User Y (*m* hosts). Sender splits steganogram into *k* parts and sends them using steganographic methods that are available in different hosts-based overt channels. User Y receives *k* pieces of the steganogram and merges them.

In the presented example maximum number of parts (*k*) into which steganogram is divided, equals a number of the different host-based overt channels – *n·m*. However, user may not want to use some hosts for hidden communication. Besides, it is worth noting that sender and receiver of the steganogram do not have to have physical access to his/her hosts. He/she can use single computer to manage other hosts.

*3) Flows and hosts-based scattering*

Hybrid SGS techniques utilize both hosts- and flows-based scattering. It means that there can be more than one flow between each pair of the sender and receiver hosts. It causes significant increase in number of available flows.

It is worth noting that not every flow must be used for steganographic purposes. Moreover, some parts of the steganogram can be redundant and can be sent using different flows. It increases resilience and the chance of the successful steganogram reception, even if some parts are lost or removed.

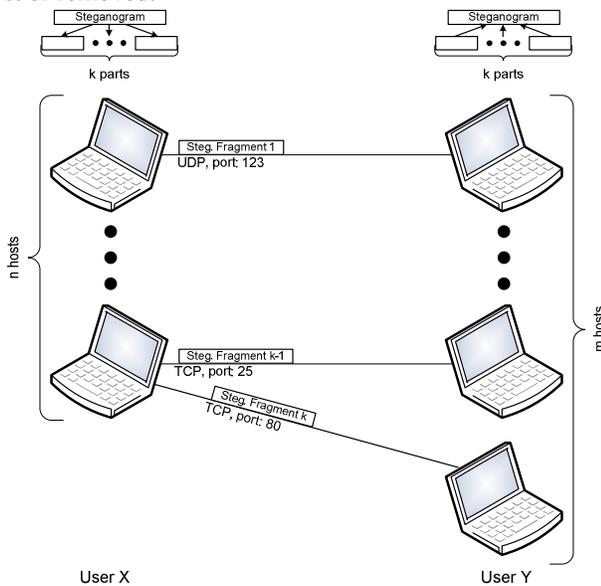

**Fig. 4 Host-based Steganogram Scattering (SGS)**

*B. Examples*

The most obvious application of the Steganogram Scattering technique is communication between two users who want to transmit secret information between them. There are many possible ways to apply different network steganography methods to achieve this goal. For example, users can use VoIP conversation as an overt channel and thus utilise e.g. LACK method [8] or one user can download file from another and use few TCP specific methods. Moreover, nowadays, it is common that ordinary users possess more than one computer device. It is also possible for users to have more than one Internet connection from different ISPs. For example, one access from cable modem ISP and another one using public wireless networks. Thus, we may divide steganogram into pieces and scatter them along different devices or steganographic method in a predetermined way. All of the abovementioned facts make SGS technique available for everyone, who wants to send hidden data.

### III. STEGANOGRAM HOPPING (SGH)

Steganogram Hopping techniques utilize periodical change of the steganographic method during single hidden connection (Fig. 5). This causes the steganogram localisation change thus making it harder to detect and extract. SGH techniques concept is similar to SGS. However, the main difference is that SGH utilizes single connection and single steganogram fragment transferring in a given moment of time, while SGS experiences no such constraints.

For example, if as a hidden data carrier SCTP connection will be chosen then the following three methods can be used for steganogram transfer [9]:

- Multi-homing based steganographic method that assigns certain steganogram bits sequence for each sender and receiver IP address. Then, depending on a hidden data bits to be transferred, consecutive retransmitted SCTP chunks are sent to the proper IP addresses.
- Multi-streaming based steganographic method that assigns certain steganogram bits sequence for each SCTP stream used. Hidden communication is then performed by sending consecutive SCTP chunks to the proper SCTP streams.
- Steganographic method that utilises the number of the SCTP chunks in SCTP packet as a hidden data carrier.

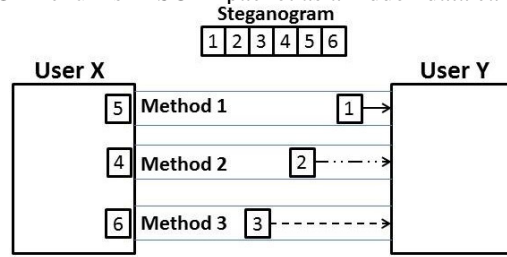

**Fig. 5 Steganogram Hopping concept**

Detection of the abovementioned methods requires statistical analysis of the SCTP connection features but for every method the analysis must be carried out differently. Analysis should be focused on communication distribution across different IP addresses or streams or on the number of the chunks in SCTP packet respectively.

Applying SGH techniques and using these three methods during single overt SCTP connection causes detection of the hidden communication and steganogram extraction to be more difficult. To discover secret data transfer, SCTP connection must be analysed in more details when compared to single SCTP steganographic method scenario.

### IV. CARRIER MODIFICATIONS CAMOUFLAGE (CMC)

DHT techniques that belong to Carrier Modifications Camouflage group are all solutions, sometimes simple,

which main aim is to mask existence of the steganogram inside the carrier. This may be achieved using a wide spectrum of actions. Examples of the CMC techniques include:

- *Intentional reduction of the steganographic bandwidth* – steganogram is inserted into the carrier less frequently, thus less secret data is sent and a chance of detection is decreased. It is the simplest way of the steganogram insertion camouflage and can be utilized for every steganographic method.
- *Adjusting the way steganographic method works during steganograms exchange* – some methods allow to influence their behaviour while transmitting steganograms to minimize the chance of disclosure. For example in Cloak method [6] one can change its parameters: number of packets $N$ and TCP flows $X$ during steganographic communication which makes detection harder to perform.
- *Adjusting steganographic method behaviour to typical users/services traffic patterns* – traffic generated by network steganography method should fit into typical traffic patterns that are present in the network and are generated by user/services etc. It involves adjusting sending of the steganograms to specific times of the day, particular frequency of the traffic generation etc. For example, previously mentioned LACK method [8] utilizes IP telephony calls to send steganograms. That is why, usage of this method should depend on the frequency, length and time of the day of the calls performed in particular network etc.
- *Utilization of the traffic characteristic features to camouflage hidden communication* – for certain steganographic methods it is possible to utilize traffic features like anomalies, level of the network parameters like number of lost packets etc. or other events to mask steganographic exchange. For example, RSTEG (Retransmission Steganography) [7] enables hidden communication by not acknowledging a successfully received packet in order to intentionally invoke retransmission. The retransmitted packet carries a steganogram instead of user data in the payload field. For RSTEG it is vital to observe network TCP retransmissions level and adjust intentional retransmissions to ones observed in the network. Another example is PadSteg (Padding Steganography) method [3] that to carry steganogram takes advantage of non-zeros frame padding that is result of the ambiguous standardisation. Existence of such unexpected anomaly masks PadSteg functioning.

## V. INTER-PROTOCOL STEGANOGRAPHY (IPS)

Typical network steganography method uses modification of a single network protocol. The protocol modification may be applied to the PDU (Protocol Data Unit), time relations between exchanged PDUs, or both (hybrid methods). This kind of network steganography is called *intra-protocol* steganography*.* The concept of the *inter-protocol* steganography (IPS) was originally introduced by Jankowski, Mazurczyk and Szczypiorski in [3]. It is defined as usage of the relationships between two or more different network protocols to enable secret communication. Protocols used by inter-protocol steganography can belong to the same layer of the e.g. TCP/IP stack or to different ones (Fig. 6). In this approach utilization of more than one protocol to enable hidden communication provides greater undetectability thus limiting the chance of disclosure. That is why one may state that IPS hides better what has already been hidden.

The first example of the inter-protocol steganography was introduced in [3] where authors proposed new steganographic system PadSteg (Padding Steganography). It benefits from utilizing relationships between Ethernet (IEEE 802.3), ARP, TCP and other protocols.

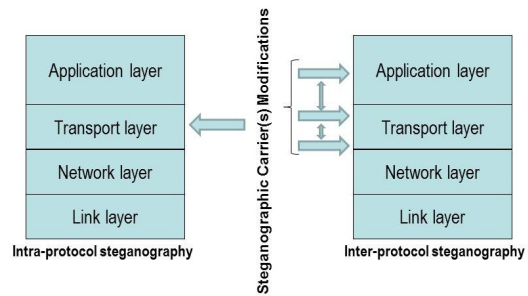

**Fig. 6 Intra- and Inter-protocol steganography comparison**

PadSteg enables secret communication in a hidden group in a LAN environment. In such group, each host willing to exchange steganograms should be able to locate and identify other hidden hosts. To provide this functionality in PadSteg, ARP protocol together with improper Ethernet frame padding are used. To exchange steganograms, improper Ethernet frame padding is utilized which is a result of using upper layer protocols like TCP, UDP, ARP or ICMP (or other network protocols that cause Ethernet frames to be padded). These protocols are called carrier-protocols as they enable transfer of the steganograms throughout the network. In other words, steganogram is inserted into Ethernet frame padding but one must always "look" at the other layer protocol (e.g. ARP, TCP, ICMP or UDP) to determine whether it contains secret data or not.

Moreover, while the secret communication takes place, hidden nodes can switch between carrier-protocols to minimize the risk of disclosure. Such mechanism is called *carrier-protocol hopping* (Fig. 7).

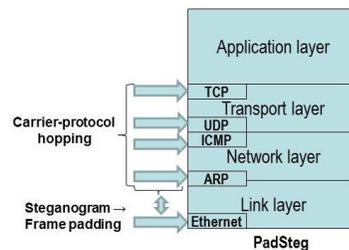

**Fig. 7 Intra- and Inter-protocol steganography comparison**

So how inter-protocol steganography can be used for typical steganographic method that utilizes single network protocol to make it more hidden? One should look for relationships between particular protocol that this method is based on and other protocols that cooperate with it. And then use this relationship to make steganographic communication harder to detect. For example, in PadSteg short frame padding is utilized which can be used as a standalone steganographic method, but by adding carrier-protocol hopping mechanism detection is encumbered.

## VI. MULTI-LEVEL STEGANOGRAPHY (MLS)

Multi-Level Steganography is a new concept of information hiding in telecommunication networks that was introduced to network steganography by Frączek, Mazurczyk and Szczypiorski in [10]. It uses features of an existing steganographic method (the upper-level method) to create a new one (the lower-level method) MLS idea is presented in Fig. 8.

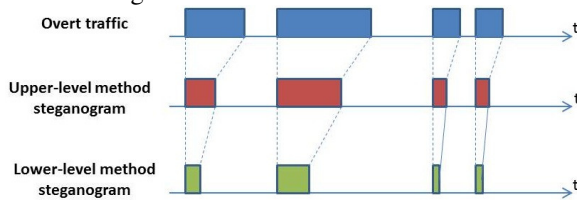

**Fig. 8 Steganographic bandwidth relationship between overt traffic and upper- and lower-level method in MLS**

MLS is based on at least two steganographic methods. First, the upper-level method uses overt traffic as a secret data carrier. The second, the lower-level method, uses the way the upper-level method operates as a carrier. The indirect carriers for lower-level methods are still packets from overt communication, but the direct carrier is another (upper-level) method

Creation of different-levels steganographic methods has two important features. Firstly, the bandwidth of the lower-level method is usually a fracture of bandwidth of the upper-level method. It is similar to relationship of overt communication bandwidth and upper-level steganography bandwidth. Secondly, the lower-level steganography is harder to detect than upper-level steganography. It results from fact that the lower-level method entirely depends on upper-level one and adversary has to detect upper-level method before he/she starts to seek for lower-level one. This is one of the reasons why MLS is a deep hiding technique. There are also other applications of MLS that allow making hidden more hidden [10], e.g. changing parameters of upper-level method.

Potentially, each known steganographic method can be used as upper-level method. The main problem is to find suitable lower-level method that will cooperate with the upper-level one. If we consider hybrid methods like LACK [8], RSTEG [7] or SCTP hybrid method [9] as upper-level method, then the lower-level method can hide bits in number of packets with steganogram of upper-level method send in established time periods. For example, if packet with steganogram of SCTP hybrid method is sent during one second period – it means binary "1", else – it means binary "0".

## VII. CONCLUSIONS AND FUTURE WORK

In this paper we introduced Deep Hiding Techniques that are defined as general techniques that can be applied to typical network steganography method to improve its undetectability and make steganogram extraction harder to perform. To authors' best knowledge presented approach is the first attempt in the state of the art to systematically describe solutions that can make steganographic communication and steganogram itself harder to detect. Knowledge of such techniques is vital when designing and developing steganalysis tools and devices. Thus, future work will be focused on analysis of each DHTs group and proposing potential detection methods.

## REFERENCES

[1] Szczypiorski K., Steganography in TCP/IP Networks. State of the Art and a Proposal of a New System – HICCUPS, Institute of Telecommunications' seminar, Warsaw University of Technology, Poland, November 2003
URL:http://krzysiek.tele.pw.edu.pl/pdf/steg-seminar-2003.pdf

[2] B. Jankowski, W. Mazurczyk, K. Szczypiorski - PadSteg: Introducing Inter-Protocol Steganography - In: Telecommunication Systems: Modelling, Analysis, Design and Management, to be published in 2013, ISSN: 1018-4864, Springer US, Journal no. 11235

[3] B. Jankowski, W. Mazurczyk, K. Szczypiorski, Information Hiding Using Improper Frame Padding, In Proc. of 14th International Telecommunications Network Strategy and Planning Symposium (Networks 2010), 27-30.09.2010, Warsaw, Poland

[4] S. Zander, G. Armitage, P. Branch, "A Survey of Covert Channels and Countermeasures in Computer Network Protocols", IEEE Communications Surveys & Tutorials, 3rd Quarter 2007, Volume: 9, Issue: 3, pp. 44-57, ISSN: 1553-877X

[5] S. Burnett, N. Feamster, S. Vempala, Chipping Away at Censorship with User-Generated Content, USENIX Security Symposium, August 2010

[6] Xiapu Luo, Edmond W. W. Chan, and Rocky K. C. Chang, Cloak: A Ten-fold Way for Reliable Covert Communications, in Proc. 12th European Symposium on Research in Computer Security (ESORICS), September 2007

[7] W. Mazurczyk, M. Smolarczyk, K. Szczypiorski, Retransmission steganography and its detection, Soft Computing, ISSN: 1432-7643 (print version), ISSN: 1433-7479 (electronic version), Journal no. 500 Springer, November 2009

[8] Mazurczyk ,W., Szczypiorski, K., Steganography of VoIP Streams, In: R. Meersman and Z. Tari (Eds.): OTM 2008, Part II - Lecture Notes in Computer Science (LNCS) 5332, Springer-Verlag Berlin Heidelberg, Proc. of The 3rd International Symposium on Information Security (IS'08), Monterrey, Mexico, November 10-11, 2008, pp. 1001-1018

[9] W. Frączek, W. Mazurczyk, K. Szczypiorski, Stream Control Transmission Protocol Steganography, The 2010 International Conference on Multimedia Information Networking and Security (MINES 2010), Nanjing, China, November 4-6, 2010

[10] W. Frączek, W. Mazurczyk, K. Szczypiorski - Multi-Level Steganography Applied to Networks - In: Proc. of: 2011 International Conference on Telecommunication Systems, Modeling and Analysis (ICTSM2011), Prague, Czech Republic, May 2011